\documentclass[fleqn,12pt,twoside]{article}
\usepackage{espcrc1}

\usepackage{graphicx}
\usepackage[figuresright]{rotating}

\title{\large Nuclear alpha-clustering, superdeformation, and
quasimolecular resonances} 

\author{C. Beck\address{Institut de Recherches Subatomiques, UMR7500,
 IN2P3-CNRS/Universit\'e Louis Pasteur, B.P. 28, F-67037
Strasbourg Cedex 2, France}}

\begin{document}
\maketitle

\begin{abstract}

{\small Nuclear alpha-clustering has been the subject of intense
study since the advent of heavy-ion accelerators. Looking back for
more than 40 years we are able today to see the connection between
quasimolecular resonances in heavy-ion collisions and extremely
deformed states in light nuclei. For example superdeformed bands have
been recently discovered in light N=Z nuclei such as $^{36}$Ar,
$^{40}$Ca, $^{48}$Cr, and $^{56}$Ni by $\gamma$-ray spectroscopy. The
search for strongly deformed shapes in N=Z nuclei is also the
domain of charged-particle spectroscopy, and our experimental group
at IReS Strasbourg has studied a number of these nuclei with the
charged particle multidetector array {\sc Icare} at the {\sc
Vivitron} Tandem facility in a systematical manner. Recently the
search for $\gamma$-decays in $^{24}$Mg has been undertaken in a
range of excitation energies where previously nuclear molecular
resonances were found in $^{12}$C+$^{12}$C collisions. The breakup
reaction $^{24}$Mg$+^{12}$C has been investigated at
E$_{lab}$($^{24}$Mg) = 130 MeV, an energy which corresponds to the
appropriate excitation energy in $^{24}$Mg for which the
$^{12}$C+$^{12}$C resonance could be related to the breakup
resonance. Very exclusive data were collected with the Binary
Reaction Spectrometer in coincidence with {\sc Euroball IV} installed
at the {\sc Vivitron}. Preliminary results on the population of
specific structures of large deformation in binary reactions will be
presented and their $\gamma$-decay determined.} 

\end{abstract}

\section{Introduction}
\label{intro}

The region of the nuclear chart between the Mg and Ni isotopes is
important for the next generation of radioactive beam experiments and
for several astrophysical applications. The interest in the
theoretical calculations for {\it sd}-shell nuclei around 
$^{40}$Ca~\cite{Caurier95,Juodagalvis00,Long01,Tanaka01,Inakura02,Sakuda02,Kanada02}
has increased recently with the discovery of highly deformed shapes
and superdeformed (SD) rotational bands in the N=Z nuclei
$^{36}$Ar~\cite{Svensson00,Svensson01},
$^{40}$Ca~\cite{Ideguchi01,Chiara03}, 
$^{42}$Ca~\cite{Lach03}, $^{48}$Cr~\cite{Lenzi96,Thummerer01} and
$^{56}$Ni~\cite{Rudolph99}. Therefore the A$_{CN}$ $\approx$ 30-60
mass region becomes of particular interest since quasimolecular
resonances have also been observed for these $\alpha$-like nuclei, in
particular, in the $^{28}$Si+$^{28}$Si 
reaction~\cite{Beck01,Nouicer99}. Although there is no experimental
evidence to link the SD bands with the higher lying rotational bands
formed by known quasimolecular resonances~\cite{Bromley60}, both
phenomena are believed to originate from highly deformed
configurations of these systems. The interpretation of resonant
structures observed in the excitation functions in various
combinations of light $\alpha$-cluster nuclei in the energy regime
from the barrier up to regions with excitation energies of 30-50~MeV
remains a subject of contemporary debate. In particular, in
collisions between two $^{12}{\rm C}$ nuclei, these resonances have
been interpreted in terms of nuclear molecules~\cite{Bromley60}.
However, in many cases these structures have been connected to
strongly deformed shapes and to the alpha-clustering phenomena,
predicted from the $\alpha$-cluster model~\cite{Marsh86},
Hartree-Fock calculations~\cite{Flocard84}, the Nilsson-Strutinsky
approach~\cite{Leander75}, and from a generalized liquid drop 
model~\cite{Royer02,Royer03a,Royer03b}. In this paper we will
show that the search for strongly deformed shapes in N=Z nuclei
is also the domain of charged-particle spectroscopy. In particular
the study of a number of $\alpha$-cluster nuclei with the charged
particle multidetector array {\sc Icare} at the {\sc Vivitron}
facility is presented in Sec.~2. In Sec.~3, the search for a
$^{12}$C+$^{12}$C molecule in $^{24}$Mg$^{*}$ is discussed with a
description of a recent particle-$\gamma$ experiment using {\sc
Euroball IV}. Finally, a summary and discussion of our results is added
in Sec.~4 for both topics. 

\section{Search for extremely extended shapes and clsuter emission
using charged-particle spectroscopy}

\begin{figure} 
  \begin{center}
    \includegraphics[width=0.68\textwidth]{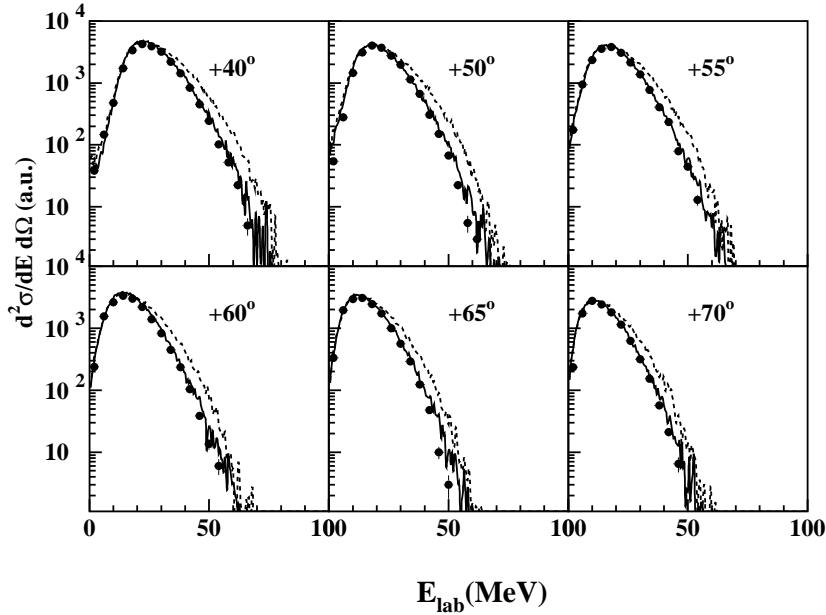}
    \caption{\small\em Exclusive energy spectra of $\alpha$ particles 
measured in the $^{28}$Si+$^{28}$Si reaction at E$_{lab}$ = 180 MeV.
Solid and dashed lines are {\sc Cacarizo} statistical-model
calculations with and without deformations effects, respectively.} 

  \end{center}
  \label{Fig.1}
\end{figure}

\noindent

Since the light charged particle (LCP) detection is relatively
simple, the analysis of their spectral shapes (see Fig.~1) can be
also considered to be a good tool for exploring nuclear deformation
and other properties of hot rotating nuclei at high angular 
momenta~\cite{Chandana02,Rousseau02} which are primarly investigated
by $\gamma$-ray spectroscopy~\cite{Styczen04}.
The LCP's emitted during the CN 
decay processes carry information on the underlying nuclear shapes
and level densities. New information on nuclear structure far above
the yrast line can be obtained from their study by a comparison with
statistical-model calculations. An experimental programme is undertaken
at the {\sc Vivitron} Tandem facility of the IReS Strasbourg
laboratory in the {\sc Icare} scattering chamber. Both the heavy
fragments (A $\geq$ 10) and their associated LCP's (protons,
deuterons, tritons, $^{3}$He, and $\alpha$ particles) were detected
in coincidence using the {\sc Icare} charged-particle multidetector
array~\cite{Chandana02,Rousseau02} which consists of nearly 50
telescopes. The properties of the LCP's emitted in the
$^{28}$Si+$^{28}$Si reaction at the bombarding energy E$_{lab}$ = 112
MeV, which corresponds to the $^{56}$Ni excitation energy of the
conjectured J$^{\pi}$ = 38$^{+}$ quasimolecular 
resonance~\cite{Beck01,Nouicer99}, were first
investigated~\cite{Chandana02}. The magnitude of the adjustments in
the yrast line position suggests deformation effects at high spin for
the $^{56}$Ni composite system in agreement with very recent
$\gamma$-ray spectroscopy data obtained at much lower
spins~\cite{Rudolph99}. This is also consistent with the generalized
liquid drop model developed by G. Royer and
coworkers~\cite{Royer03b}. The extent to which the resonant behaviour
is responsible to the observed nuclear deformation is still an open
question. To resolve this issue, we have performed a subsequent
$^{28}$Si+$^{28}$Si experiment at E$_{lab}$ = 180 MeV outside the
``molecular window'' where quasimolecular resonances are known to
disappear~\cite{Beck94}. The strong cluster emission of $^{8}$Be
which was observed by the $^{28}$Si+$^{12}$C
reaction~\cite{Rousseau02} has motivated the search for similar
effects in the $^{27}$Al+$^{12}$C~\cite{Beck03a},
$^{31}$P+$^{12}$C~\cite{Beck03a}, $^{32}$S+$^{12}$C~\cite{Papka03a},
and $^{16}$O+$^{28}$Si~\cite{Papka03b} reactions. Deformation effects
have also been investigated in great detail in
$^{28}$Si+$^{12}$C~\cite{Rousseau02}, in
$^{32}$S+$^{12}$C~\cite{Papka03a}, and in $^{16}$O+$^{28}$Si
\cite{Papka03b}. 

\begin{figure}
  \begin{center}
    \includegraphics[width=0.68\textwidth]{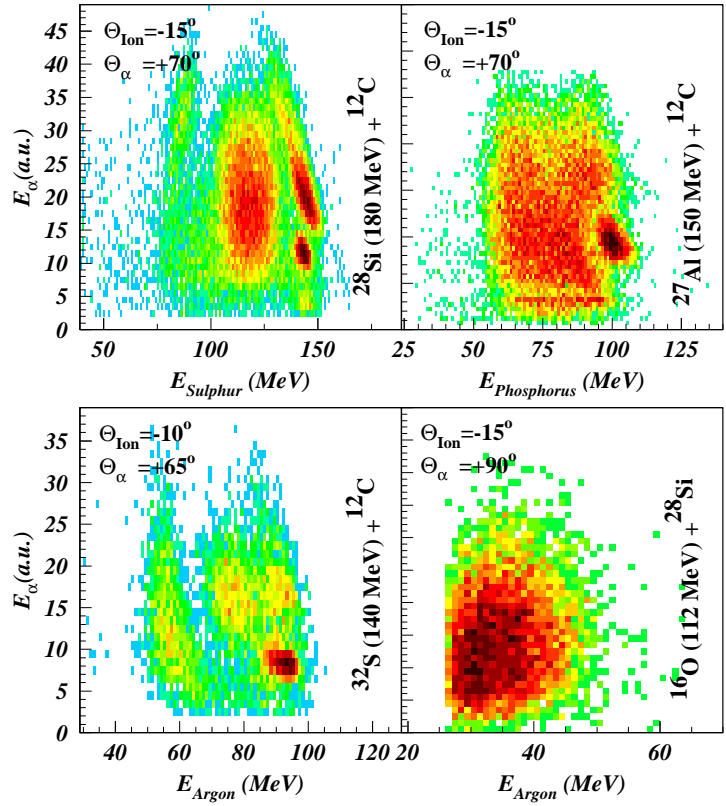}
    \caption{\small\em Energy-correlation plots between $\alpha$
particles and ER's for the $^{28}$Si+$^{12}$C, $^{27}$Al+$^{12}$C,
$^{32}$S+$^{12}$C, and $^{16}$O+$^{28}$Si reations at the indicated
angle settings.}
  \end{center}
  \label{Fig.2}
\end{figure}

Exclusive energy spectra measured for $\alpha$ particles are
displayed in Fig. 1 for $^{28}$Si+$^{28}$Si. The analysis of the data
has been performed using {\sc Cacarizo}~\cite{Chandana02}, the Monte
Carlo version of the statistical-model code {\sc Cascade}. The
parameters needed for the statistical description, i.e. the nuclear
level densities and the barrier transmission probabilities, are
usually obtained from the study of LCP evaporation spectra. The
change in the emission barriers and, correspondingly, the
transmission probabilities affects the lower energy part of the
calculated evaporation spectra. On the other hand the high-energy
part of the $\alpha$-particle spectra depends critically on the available
phase space obtained from the level densities at high spin. This is
clearly shown by the solid and dashed lines of Fig.~1 for
$^{28}$Si+$^{28}$Si at E$_{lab}$ = 180 MeV. The level density,
$\rho(E,J)$, for a given angular momentum $J$ and energy $E$ is given
by the well known Fermi gas expression: 

$\rho(E,J) = {\frac{(2J+1)}{12}}a^{1/2}
           ({\frac{ \hbar^2}{2 {\cal J}_{eff}}}) ^{3/2}
           {\frac{1}{(E-\Delta-T-E_J)^2} }exp(2[a(E-\Delta-T-E_J)]^{1/2})$ 

\noindent
where $a$ is the level density parameter set equal to $a$ = A/8
MeV$^{-1}$ (A is the mass number), T is the ``nuclear" temperature,
and $\Delta$ is the pairing correction, E$_J$ = $\frac{ \hbar^2}{2
{\cal J}_{eff}}$J(J+1) is the rotational energy, ${\cal J}_{eff}=
{\cal J}_{sphere} \times (1+\delta_1J^2+\delta_2J^4)$ is the
effective moment of inertia,  ${\cal J}_{sphere}$ =
${\frac{2}{5}}$AR$^{2}$ = ${\frac{2}{5}}$A$^{5/3}$r$_{0}^{2}$ is the
rigid body moment of inertia of a spherical nucleus with radius
parameter r$_{0}$, and $\delta_1$ and $\delta_2$ are the
deformability parameters. The solid lines in Fig.~1 show the
predictions using the same parameter set with deformation effects
that has been used at E$_{lab}$ = 180 MeV \cite{Chandana02}.
Predictions with the parameters of the finite-range liquid drop
model~\cite{Sierk86} are unable to reproduce the data (dashed lines).
Since the highest incident energy is outside of the ``molecular
window'' \cite{Beck94}, we conclude that the highly deformed shapes
(with $\beta$ $\approx$ 0.5) observed for both energies are not
related to the $^{28}$Si+$^{28}$Si quasimolecular 
resonances~\cite{Beck01,Nouicer99,Beck94}. However they may still be
linked to shape isomerism of rotating $^{56}$Ni as suggested very
recently by G. Royer {\it et al.}~\cite{Royer03b}. Similarly, the
extremely extended shapes that are needed to explain the 
$\alpha$-particle
spectra measured in the $^{28}$Si+$^{12}$C~\cite{Rousseau02},
$^{32}$S+$^{12}$C~\cite{Papka03a}, and
$^{28}$Si+$^{16}$O~\cite{Papka03b} reactions should correspond to
shape isomerism of rotating $^{40}$Ca~\cite{Royer02} and
$^{44}$Ti~\cite{Royer03a,Royer03b}. 

Superimposed on the Maxwellian shapes typical of their evaporative
origin, nonstatistical $\alpha$-particles components were found in
the energy spectra measured in coincidence with S residues in the
$^{28}$Si+$^{12}$C reaction~\cite{Rousseau02} and Ar residues for the
$^{32}$S+$^{12}$C reaction~\cite{Papka03a}, respectively. In both
reactions, these additional components attributed to the decay of
unbound $^8$Be nuclei are more easily observed in the corresponding
energy correlation plot displayed in Fig.~3 (left side of the upper
panel). They appear as well defined peaks lying outside the
``statistical evaporation region'' which is consistent with {\sc
Cacarizo} calculations (see Fig.~10 of Ref.~\cite{Rousseau02}). Their
``folding angles'' are compatible with the two-body kinematics
required for the $^{32}$S+$^{8}$Be binary exit-channel. To clearly
establish the mechanism resulting in these yields subsequent
experiments have been undertaken with other reaction with a $^{12}$C
target~\cite{Beck03a} such as $^{27}$Al(150 MeV)+$^{12}$C and
$^{31}$P(112 MeV)+$^{12}$C. Fig.~2 also displays the
energy-correlation plots between $\alpha$ particles and ER's for three of these
reactions~\cite{Beck03a} and for the $^{16}$O+$^{28}$Si reaction at
E$_{lab}$($^{16}$O) = 112 MeV~\cite{Papka03b}. The fact that the
two-body components do not show-up for this last
reaction~\cite{Papka03b} indicates the binary nature of the 
$\alpha$-particle
peaks present in the reactions involving a $^{12}$C target. This
hypothesis is consistent with the cluster-transfer picture proposed
by Morgenstern {\it et al.}~\cite{Morgenstern86} for incomplete
fusion mechanisms. 

To summarize this topic, we have confirmed that the charged-particle
spectroscopy appears to be a very efficient technique to search for
superdeformed and hyperdeformed shapes in light ions, as well as to
identify $^{8}$Be and $^{12}$C $\alpha$-cluster emissions with excellent
selectivity at high excitation energy. 

\section{Search for the $^{12}$C+$^{12}$C molecule in the
$^{24}$Mg+$^{12}$C breakup reaction}

In the presentation of the second topic of this paper we investigate
the question whether $^{12}$C+$^{12}$C molecular resonances represent
true cluster states in the $^{24}$Mg compound system, or whether they
simply reflect scattering states in the ion-ion potential is still
unresolved. Various decay branches from the highly excited
$^{24}$Mg$^*$ nucleus, including the emission of $\alpha$ particles
or heavier fragments such as $^{8}$Be and $^{12}$C, are possibly
available. However, $\gamma$-decays have not been observed so far.
Actually the $\gamma$-ray branches are predicted to be rather small
at these excitation energies, although some experiments have been
reported~\cite{McGrath81,Metag82,Haas97}, which have searched for
these very small branches expected in the range of
$10^{-4}~-~10^{-5}$~fractions of the total width~\cite{Uegaki98,Beck03b}. The
rotational bands built on the knowledge of the measured spins and
excitation energies can be extended to rather small angular momenta,
where finally the $\gamma$-decay becomes a larger part of the total
width. The population of such states in $\alpha$-cluster nuclei,
which are lying below the threshold for fission decays and for other
particle decays, is favored in binary reactions, where at a fixed
incident energy the composite nucleus is formed with an excitation
energy range governed by the two-body reaction kinematics. These
states may be coupled to intrinsic states of $^{24}$Mg$^{*}$ as
populated by a breakup process (via resonances) as shown in previous
works~\cite{Fulton86,Curtis95,Singer00}. The $^{24}$Mg+$^{12}$C
reaction has been extensively investigated by several measurements of
the $^{12}$C($^{24}$Mg,$^{12}$C$^{12}$C)$^{12}$C breakup
channel~\cite{Fulton86,Curtis95,Singer00}. Sequential breakups are
found to occur from specific states in $^{24}$Mg at excitation
energies ranging from 20 to 35 MeV, which are linked to the ground
state and also have an appreciable overlap with the $^{12}$C+$^{12}$C
quasi-molecular configuration. Several attempts \cite{Curtis95} were
made to link the $^{12}$C+$^{12}$C barrier resonances
\cite{Bromley60} with the breakup states. The underlying reaction
mechanism is now fairly well established~\cite{Singer00} and many of
the barrier resonances appear to be correlated indicating that a
common structure may exist in both instances. This is another
indication of the possible link between barrier resonances and
secondary minima in the compound nucleus~\cite{Leander75}. 

The study of particle-$\gamma$ coincidences in binary reactions in
reverse kinematics is probably a unique tool for the search for
extreme shapes related to clustering. In this way the
$^{24}$Mg+$^{12}$C reaction has been investigated with high
selectivity at E$_{lab}$($^{24}$Mg) = 130 MeV with the Binary
Reaction Spectrometer~\cite{Thummerer01} (BRS) in coincidence with
{\sc Euroball IV} installed at the {\sc Vivitron}~\cite{Beck03b}. The
choice of the $^{12}{\rm C}(^{24}{\rm Mg},^{12}{\rm C})^{24}{\rm
Mg^{*}}$ reaction implies that for an incident energy of E$_{lab}$ =
130~MeV an excitation energy range up to E$^{*}$ = 30~MeV in
$^{24}$Mg is covered~\cite{Curtis95}. The BRS gives access to a novel
approach to the study of nuclei at large deformations~\cite{Beck03b}.
The excellent channel selection capability of binary and/or ternary
fragments gives a powerful identification among the reaction
channels, implying that {\sc Euroball IV} is used mostly with one or
two-fold multiplicities, for which the total $\gamma$-ray efficiency
is very high. The BRS trigger consists of a kinematical coincidence
set-up combining two large-area heavy-ion telescopes. Both detector
telescopes comprise each a two-dimensional position sensitive
low-pressure multiwire chamber in conjunction with a Bragg-curve
ionization chamber. All detection planes are four-fold subdivided in
order to improve the resolution and to increase the counting rate
capability (100 k-events/s). The two-body Q-value has been
reconstructed using events for which both fragments are in well
selected states chosen for spectroscopy purposes as well as to
determine the reaction mechanism responsible for the population of
these peculiar states. Fig.~3 displays a typical example of a
two-dimensional Bragg-Peak versus energy spectrum obtained for the
$^{24}$Mg+$^{12}$C reaction. This coincident spectrum shows the
excellent charge discrimination achieved with the Bragg-curve
ionization chambers. The Z=12 gate, which is shown in Fig.~3, will be
used in the following for the processing of the $\gamma$-ray spectra
of the $^{24}$Mg nucleus of interest.

The inverse kinematics of the $^{24}$Mg+$^{12}$C reaction and the
negative Q-values give ideal conditions for the trigger on the BRS,
because the angular range is optimum for $\theta_lab$ =
12$^\circ$-40$^\circ$ in the lab-system (with $\theta_lab$ =
12$^\circ$-25$^\circ$ for the recoils) and because the solid angle
transformation gives a factor~10 for the detection of the heavy
fragments. Thus we have been able to cover a large part of the
angular distribution of the binary process with high efficiency, and
a selection of events in particular angular ranges has been achieved.
In binary exit-channels the exclusive detection of both ejectiles
allows precise Q-value determination, Z-resolution and simultaneously
optimal Doppler-shift correction. Fig.~4 displays a Doppler-corrected
$\gamma$-ray spectrum for $^{24}$Mg events in coincidence with the
Z=12 gate defined in the Bragg-Peak vs energy spectrum of Fig.~3. All
known transitions of $^{24}$Mg~\cite{Beck01,Wiedenhover01} can be
identified in the energy range depicted. As expected we see decays
feeding the yrast line of $^{24}$Mg up to the 8$^{+}_{2}$ level. The
population of some of the observed states, in particular, the
2$^{+}$, 3$^{+}$ and 4$^{+}$ members of the K=2 rotational band,
appears to be selectively enhanced. The strong population of the K=2
band has also been observed in the $^{12}$C($^{12}$C,$\gamma$)
radiative capture reaction~\cite{Jenkins03}. Furthermore, there is an
indication of a $\gamma$-ray around 5.95 MeV which may be identified
with the 10$^{+}_{1}$ $\rightarrow$ 8$^{+}_{2}$ transition as
proposed in Ref.~\cite{Wiedenhover01}. It has been checked in the
$\gamma$-$\gamma$ coincidences that most of the states of Fig.~4
belong to cascades which contain the characteristic 1368 keV
$\gamma$-ray and pass through the lowest 2$^{+}$ state in $^{24}$Mg.
Still a number of transitions in the high-energy part of the spectrum
(6000~keV~-~8000~keV) have not been clearly identified. The reason
why the search for a $\gamma$-decay in $^{12}$C+$^{12}$C has not been
conclusive so far~\cite{McGrath81,Metag82,Haas97} is due to the
excitation energy in $^{24}$Mg as well as the spin region
(8$\hbar$-12$\hbar$) which were chosen too high. The next step of the
analysis will be the use of the BRS trigger in order to select the
excitation energy range by the two-body Q-value (in the
$^{12}$C+$^{24}$Mg channel), and thus we will be able to study the
region around the decay barriers, where $\gamma$-decay becomes
observable. According to recent predictions $\gamma$-rays from
6$^{+}$ $\rightarrow$ 4$^{+}$ should have measurable branching
ratios. Work is currently in progress to analyse the $\gamma$ rays
from the $^{12}$C($^{24}$Mg,$^{12}$C $^{12}$C)$^{12}$C ternary
breakup reaction. 

\begin{figure}
  \begin{center}
    \includegraphics[width=0.68\textwidth]{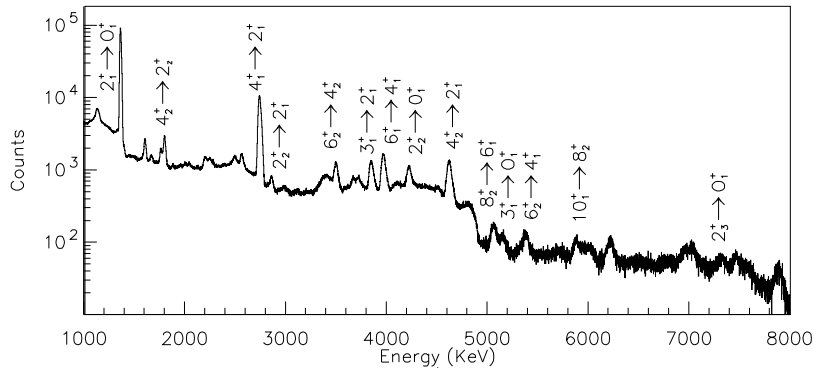}
    \caption{\small\em Doppler corrected $\gamma$-ray spectrum for
                       $^{24}$Mg,
                       using particle-particle-$\gamma$
                       coincidences, measured
                       in the $^{24}$Mg(130 MeV)+$^{12}$C reaction
                       with the BRS/{\sc Euroball IV} detection system 
(see text). }
  \end{center}
  \label{Fig.4}
\end{figure}

\section{Summary and conclusions}
\label{concl}

The occurence of highly deformed configurations in light N=Z nuclei
and their possible link with alpha-clustering have been investigated
at the {\sc Vivitron} Tandem facility of the IReS Strasbourg by using
two complementary experimental techniques: either the {\sc Icare}
charged-particle multidetector array or the BRS/{\sc Euroball IV}
detection system. In the first case. the properties of the emitted
LCP's in several reactions have been analysed with the {\sc Cacarizo}
statistical-model code that was adapted to calculate evaporation
spectra and angular distributions for deformed nuclei. The measured
observables such as energy spectra in-plane and out-of-plane angular
correlations are well described by Hauser-Feshbach calculations which
include spin-dependent level densities. The magnitude of the
adjustements in the yrast line suggests deformations at high spins
that are far in excess of those predicted by the finite-range liquid
drop model~\cite{Sierk86}. The deformation parameters deduced for the
$^{28}$Si+$^{12}$C~\cite{Rousseau02},
$^{32}$S+$^{12}$C~\cite{Papka03a},
$^{16}$O+$^{28}$Si~\cite{Papka03b}, and
$^{28}$Si+$^{28}$Si~\cite{Chandana02} reactions are comparable to
recent $\gamma$-ray spectroscopy data for the $^{40}$Ca
nucleus~\cite{Ideguchi01} and for the $^{56}$Ni
nucleus~\cite{Rudolph99} at much lower spins. The use of large
$\gamma$-ray multidetector arrays will be helpful to extend the
existing level scheme of the $^{44}$Ti nucleus~\cite{Leary2000} for
the search for weakly populated SD rotational bands (predicted by the
theory~\cite{Horiuchi03}) equivalent to those discovered in
$^{40}$Ca~\cite{Ideguchi01}. For the both the $^{28}$Si+$^{12}$C and
$^{32}$Si+$^{12}$C reactions the components which are found in the
$\alpha$-particle energy spectra measured in coincidence with S and
Ar residues are attributed to the cluster decay of unbound $^8$Be
nuclei~\cite{Rousseau02,Papka03a}. The hypothesis of the binary
nature of the $^{8}$Be cluster emission is consistent with the
preliminary results found for the two other reactions involving a
$^{12}$C target~\cite{Beck03a}: $^{27}$Al+$^{12}$C at E$_{lab}$ = 150
MeV, $^{31}$P+$^{12}$C at E$_{lab}$ = 112 MeV and 220 MeV. In the
second topic of the paper the link of alpha-clustering and
quasimolecular resonances has been discussed with the search for the
$^{12}$C+$^{12}$C molecule populated by the $^{24}$Mg+$^{12}$C
breakup reaction. The most intriguing result is the strong population
of the K=2 band of the $^{24}$Mg nucleus that has also been observed
in an exploratory investigation of the $^{12}$C($^{12}$C,$\gamma$)
radiative capture reaction~\cite{Jenkins03}. New experiments are
planned in the near future with highly efficient spectrometers
(the {\sc Dragon separator} at {\sc Triumf} and the {\sc Fma} at
Argonne) to study this possible overlap of the $^{24}$Mg states
observed in the present work with radiative capture states more in
detail. As far as the $\gamma$-ray spectroscopy is concerned, the
coexistence of $\alpha$-cluster states and superdeformed states
predicted in $^{32}$S by antisymmetrized molecular dynamics (AMD)
studies~\cite{Horiuchi03} is still an experimental challenge. \\ 

\noindent 
{\small
{\bf Acknowledgments:} I would like to acknowledge the physicists
of {\sc Icare} and BRS/{\sc Euroball IV} collaborations, with
special thanks to M. Rousseau, P. Papka, A. S\`anchez i Zafra, C.
Bhattacharya, and S. Thummerer. We thank the staff of the {\sc
Vivitron} for providing us with good stable beams, M.A. Saettel for
preparing targets, and J. Devin and C. Fuchs for their excellent
support during the experiments. This work was supported by the french
IN2P3/CNRS and the EC Euroviv contract HPRI-CT-1999-00078.}

\end{document}